\pdfoutput=1

\documentclass[11pt]{article}

\usepackage[]{acl}

\usepackage{times}
\usepackage{latexsym}

\usepackage[T1]{fontenc}

\usepackage[utf8]{inputenc}

\usepackage{microtype}

\usepackage{inconsolata}

\usepackage{graphicx}

\usepackage{multirow}
\usepackage{tcolorbox}
\usepackage{tabularx}
\usepackage{makecell}
\usepackage{amsmath,amssymb}
\usepackage{physics}
\usepackage{booktabs}
\usepackage{enumitem}
\usepackage{subcaption}
\usepackage[switch]{lineno}
\usepackage{listings}
\usepackage{xcolor}
\usepackage{authblk}
\usepackage{fontawesome}
\usepackage{cleveref}
\usepackage{tikz}

\lstdefinelanguage{json}{
    basicstyle=\ttfamily\scriptsize,
    numberstyle=\tiny\color{gray},
    stepnumber=1,
    numbersep=8pt,
    showstringspaces=false,
    breaklines=true,
    frame=single,
    backgroundcolor=\color{white},
    literate=
     *{0}{{{\color{blue}0}}}{1}
      {1}{{{\color{blue}1}}}{1}
      {2}{{{\color{blue}2}}}{1}
      {3}{{{\color{blue}3}}}{1}
      {4}{{{\color{blue}4}}}{1}
      {5}{{{\color{blue}5}}}{1}
      {6}{{{\color{blue}6}}}{1}
      {7}{{{\color{blue}7}}}{1}
      {8}{{{\color{blue}8}}}{1}
      {9}{{{\color{blue}9}}}{1}
      {:}{{{\color{blue}:}}}{1}
      {"}{{{\color{blue}"}}}{1}
      {,}{{{\color{blue},}}}{1}
      {\{}{{{\color{blue}\{}}}{1}
      {\}}{{{\color{blue}\}}}}{1}
      {[}{{{\color{blue}[}}}{1}
      {]}{{{\color{blue}]}}}{1},
    stringstyle=\color{green},
    keywordstyle=\color{red}\bfseries
}

\newcommand{\highlight}[1]{\colorbox{yellow}{#1}}
\newcommand{\tool} {{\tt ConCodeEval}}

\title{ConCodeEval: Evaluating Large Language Models for Code Constraints in Domain-Specific Languages}

\author[1]{\bf Mehant Kammakomati$^*$}
\author[2]{\bf Sameer Pimparkhede$^*$}
\author[1]{\bf Srikanth G. Tamilselvam} 
\author[1]{\\ \bf Prince Kumar}
\author[2]{\bf Pushpak Bhattacharyya}

\affil[1]{IBM Research}
\affil[2]{IIT Bombay}
\affil[1]{{\{mehant.kammakomati2,prince.kumar12\}@ibm.com}}
\affil[1]{{\{srikanth.tamilselvam\}@in.ibm.com}}
\affil[2]{\{sameerp,pb\}@cse.iitb.ac.in}

\begin{document}
\maketitle
\def\thefootnote{\arabic{footnote}}

\begin{abstract}
System-level programming is essential for modern enterprise infrastructure, enabling the automation and management of complex systems through declarative code. Developers write this code based on schemas, which themselves are a form of code that defines constraints like data types and required fields. These schemas help ensure operational correctness and smooth integration across systems. However, as enterprise schemas become complex, manually writing code adhering to these constraints becomes challenging for developers. Large Language Models (LLMs) have demonstrated potential in code generation and natural language understanding, particularly in zero-shot and few-shot settings. However, applying LLMs to handle constraints represented in code, essential for system-level programming rather than natural language, has not been explored. Hence, we introduce \tool{}, a study across two key dimensions: format and constraint efficacy with a first-of-its-kind benchmark involving two novel experiments for code constraints across five representations (JSON, YAML, XML, Python, and natural language). Our findings suggest that conscious use of representations can lead to optimal use of LLMs in enterprise use cases involving constraints. Nonetheless, LLMs still struggle with code constraints motivating the need for innovation in this direction.

\end{abstract}

\section{Introduction}

System-level programming is the backbone of modern enterprise infrastructure, enabling developers to define, manage, and automate complex systems seamlessly. Numerous enterprises use concepts like Infrastructure as Code\footnote{\url{https://en.wikipedia.org/wiki/Infrastructure_as_code}} (IaC) to let developers write declarative code. Such code must adhere to constraints called schemas, which define rules, including data types, required fields, and valid value ranges, ensuring operational correctness and smooth integration. For instance, the schema in Listing \ref{lst:task1} mandates an array of even numbers within specific bounds, containing 1 to 7 elements. 

\begin{lstlisting}[language=json, caption=The JSON sample generated (highlighted in yellow) by the Granite 20B model does not adhere to the \emph{minContains} and subsequent numerical constraints specified in the schema., label={lst:task1}, escapechar=@]

Write a JSON sample with field values as per the JSON format schema given below.

{
    "type": "array",
    "contains": {
        "type": "number",
        "multipleOf": 2,
        "exclusiveMinimum": 0,
        "exclusiveMaximum": 65535
    },
    "minContains": 1,
    "maxContains": 7
}

JSON sample:
```
@\highlight{[2, 3, 4, 6, 8, 10, 12, 14]}@
@\highlight{```}@
\end{lstlisting}

Schemas are crucial in real-world enterprise settings. For instance, deploying a database service in an OpenShift cluster involves writing compliant code with the correct attributes, such as the number of instances, port number to expose, compute to allocate, etc.
Developers write system-level code in structured Domain Specific Languages (DSLs) such as JSON, YAML, XML, or Python, adhering to strict schema constraints. However, enterprise schemas are often complex and difficult to learn, slowing development and increasing errors.
As a result, the need for automated and accurate systems for system-level programming is increasing leading to products such as Ansible Lightspeed \cite{Lightspeed}.




LLMs have shown great promise in generating coherent text and code in zero-shot and few-shot settings, making them highly appealing for system-level coding \citep{NEURIPS2020_1457c0d6, roziere2023code, mishra2024granite}. 
Using LLMs to handle constraints represented in natural language (NL) has been extensively explored for tasks like poem generation and summarization \cite{sun-etal-2023-evaluating}. However, Unlike these natural language tasks, constraints are often represented as code for system-level programming; hence, evaluating LLMs requires a different approach. In addition to assessing how well models adhere to constraints expressed in natural language, we must examine their ability to process, interpret, and generate structured formats while ensuring schema compliance. To ensure this, we evaluate LLMs under two key dimensions: \textbf{Format Efficacy} and \textbf{Constraint Efficacy}. 

Format efficacy involves studying the performance of LLMs on varying constraint representations that form the input and output representations downstream enterprise use cases can consume. Specifically, we aim to answer the following research questions (RQ) for format efficacy: 1) Which format is optimally suited for constraint and output representation? 2) What is the trade-off between performance and context length cost? While constraint efficacy involves studying LLMs' performance on various schema constraints within a format. Precisely, we aim to answer the following research questions related to constraint efficacy: 1) How does performance vary across different types of constraints? 2) What are the ideal positions for constraints in the schema for better adherence?



We prepare first-of-its-kind benchmark test set and conduct two experiments involving $5$ schema formats (JSON, YAML, XML, Python, and NL) and $3$ output formats (JSON, YAML, and XML) resulting $15$ combinations of use cases to investigate the aforementioned research questions. 1) Data as Code Generation (Section \ref{exp1desc}. 2) Data Validation (Section \ref{exp2desc}). This study provides insights into leveraging LLMs effectively for system-level programming tasks involving code constraints in enterprises.

\textbf{Our contributions are:}

\begin{enumerate}
    \item First-of-its-kind study of language models for crucial industry use case of system-level programming involving code format constraints across four key dimensions: Format and Constraint efficacy.
    \item A benchmark test set consisting of 602 schema samples, each containing multiple instructions. Each schema sample in our test set is represented in 5 different language formats (JSON, YAML, XML, Python, and NL).
    \item Comparative and qualitative analysis of state-of-the-art language models involving code generation from fine-grained schema instructions and code validation against schemas. To the best of our knowledge, we are the first to evaluate LLMs code constraint competency.
\end{enumerate}

\section{Experiments}

\subsection{Data as Code Generation in DSL}\label{task1para}

\begin{figure}[ht]
 \begin{center}
 \includegraphics[width=0.4\textwidth]
 {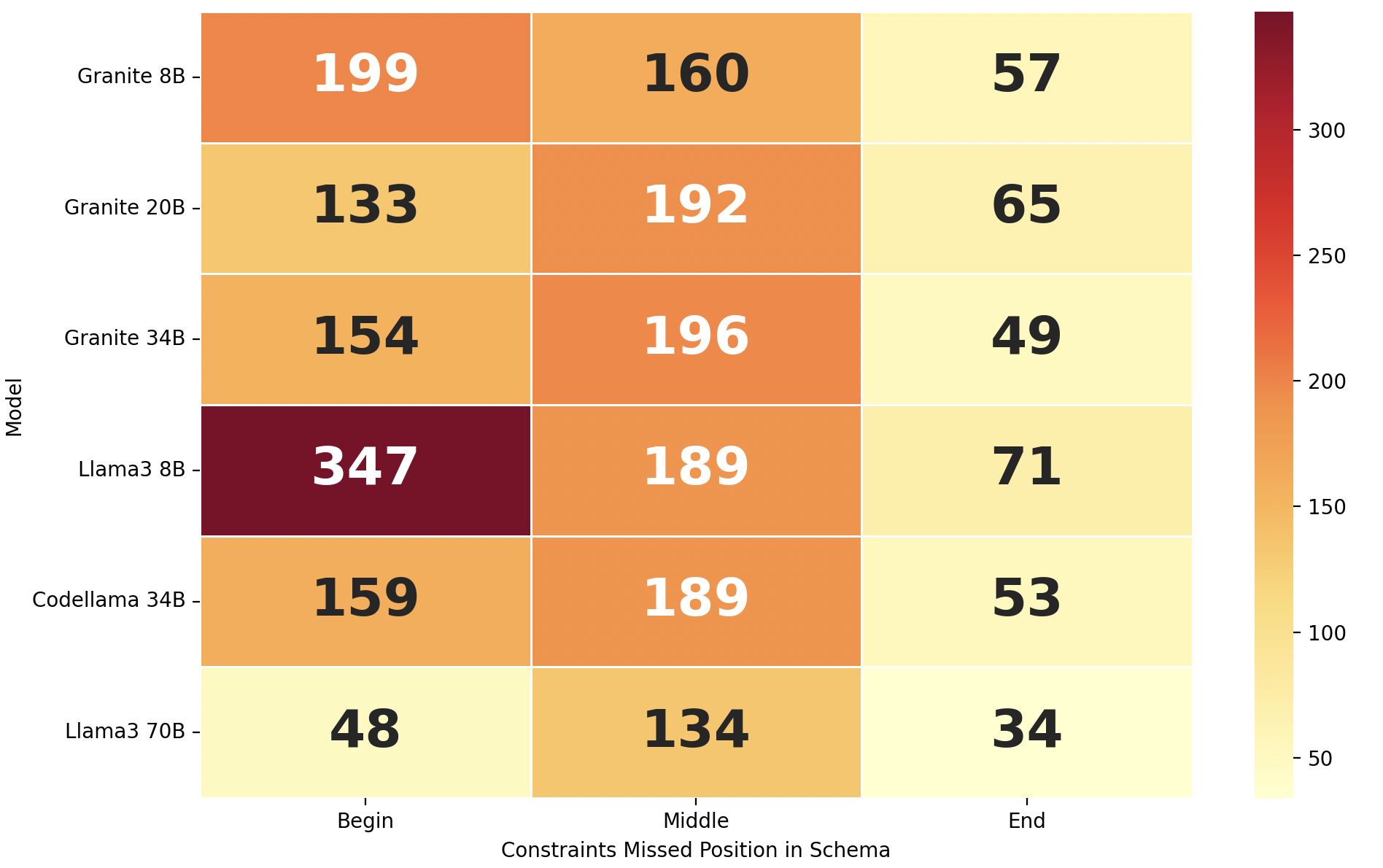}
 \end{center}
\caption{Uniform trend of steep decline in performance across models for constraints positioned in the middle and beginning of the JSON schema context and output for data as code generation experiment. We divide the schema into $3$ portions, Begin, Middle, and End, and put the violated constraints based on their locality into either of these three buckets.}
\label{needle figure}
\end{figure}

\begin{table*}[!ht]
\setlength\tabcolsep{14pt}
\scriptsize
\centering
\begin{tabularx}{\textwidth}{c|c|cccccc}
\toprule
 &
   &
  \multicolumn{6}{c}{Output Representation} \\ \midrule
 &
   &
  \multicolumn{2}{c|}{JSON} &
  \multicolumn{2}{c|}{YAML} &
  \multicolumn{2}{c}{XML} \\ \midrule
Model &
  Schema &
  \multicolumn{1}{c|}{Gen Acc} &
  \multicolumn{1}{c|}{Val Acc} &
  \multicolumn{1}{c|}{Gen Acc} &
  \multicolumn{1}{c|}{Val Acc} &
  \multicolumn{1}{c|}{Gen Acc} &
  Val Acc \\ \midrule
Llama3 8B &
  \multirow{5}{*}{JSON} &
  \multicolumn{1}{c|}{28.2} &
  \multicolumn{1}{c|}{56.0} &
  \multicolumn{1}{c|}{29.2} &
  \multicolumn{1}{c|}{45.0} &
  \multicolumn{1}{c|}{7.9} &
  47.0 \\
Granite 8B &
   &
  \multicolumn{1}{c|}{47.5} &
  \multicolumn{1}{c|}{56.0} &
  \multicolumn{1}{c|}{24.7} &
  \multicolumn{1}{c|}{55.0} &
  \multicolumn{1}{c|}{5.1} &
  45.0 \\
Granite 20B &
   &
  \multicolumn{1}{c|}{50.4} &
  \multicolumn{1}{c|}{52.0} &
  \multicolumn{1}{c|}{37.7} &
  \multicolumn{1}{c|}{44.0} &
  \multicolumn{1}{c|}{10.1} &
  53.0 \\
Granite 34B &
   &
  \multicolumn{1}{c|}{53.3} &
  \multicolumn{1}{c|}{64.0} &
  \multicolumn{1}{c|}{32.2} &
  \multicolumn{1}{c|}{57.0} &
  \multicolumn{1}{c|}{11.2} &
  \textbf{65.0} \\
Codellama 34B &
   &
  \multicolumn{1}{c|}{58.4} &
  \multicolumn{1}{c|}{64.0} &
  \multicolumn{1}{c|}{23.0} &
  \multicolumn{1}{c|}{54.0} &
  \multicolumn{1}{c|}{9.4} &
  53.0 \\ 
  
\textbf{\faTrophy{} Llama3 70B} &
   &
  \multicolumn{1}{c|}{\textbf{62.8}} &
  \multicolumn{1}{c|}{\textbf{67.0}} &
  \multicolumn{1}{c|}{\textbf{40.1}} &
  \multicolumn{1}{c|}{\textbf{58.4}} &
  \multicolumn{1}{c|}{\textbf{18.9}} &
  55.7 \\\midrule
Llama3 8B &
  \multirow{5}{*}{XML} &
  \multicolumn{1}{c|}{10.2} &
  \multicolumn{1}{c|}{37.0} &
  \multicolumn{1}{c|}{22.5} &
  \multicolumn{1}{c|}{42.0} &
  \multicolumn{1}{c|}{10.2} &
  46.0 \\
Granite 8B &
   &
  \multicolumn{1}{c|}{18.9} &
  \multicolumn{1}{c|}{47.0} &
  \multicolumn{1}{c|}{12.1} &
  \multicolumn{1}{c|}{44.0} &
  \multicolumn{1}{c|}{8.4} &
  52.0 \\
Granite 20B &
   &
  \multicolumn{1}{c|}{24.0} &
  \multicolumn{1}{c|}{37.0} &
  \multicolumn{1}{c|}{12.4} &
  \multicolumn{1}{c|}{47.0} &
  \multicolumn{1}{c|}{8.6} &
  57.0 \\
Granite 34B &
   &
  \multicolumn{1}{c|}{18.7} &
  \multicolumn{1}{c|}{68.0} &
  \multicolumn{1}{c|}{18.1} &
  \multicolumn{1}{c|}{58.0} &
  \multicolumn{1}{c|}{8.6} &
  \textbf{58.0} \\
Codellama 34B &
   &
  \multicolumn{1}{c|}{8.8} &
  \multicolumn{1}{c|}{46.0} &
  \multicolumn{1}{c|}{14.2} &
  \multicolumn{1}{c|}{46.0} &
  \multicolumn{1}{c|}{8.6} &
  50.0 \\ 
\textbf{\faTrophy{} Llama3 70B} &
   &
  \multicolumn{1}{c|}{\textbf{28.4}} &
  \multicolumn{1}{c|}{\textbf{70.3}} &
  \multicolumn{1}{c|}{\textbf{24.8}} &
  \multicolumn{1}{c|}{\textbf{60.1}} &
  \multicolumn{1}{c|}{\textbf{16.6}} &
  54.2 \\\midrule
Llama3 8B &
  \multirow{5}{*}{YAML} &
  \multicolumn{1}{c|}{25.9} &
  \multicolumn{1}{c|}{46.0} &
  \multicolumn{1}{c|}{8.1} &
  \multicolumn{1}{c|}{44.0} &
  \multicolumn{1}{c|}{6.4} &
  45.0 \\
Granite 8B &
   &
  \multicolumn{1}{c|}{47.0} &
  \multicolumn{1}{c|}{47.0} &
  \multicolumn{1}{c|}{15.7} &
  \multicolumn{1}{c|}{50.0} &
  \multicolumn{1}{c|}{8.6} &
  44.0 \\
Granite 20B &
   &
  \multicolumn{1}{c|}{34.7} &
  \multicolumn{1}{c|}{31.0} &
  \multicolumn{1}{c|}{25.9} &
  \multicolumn{1}{c|}{38.0} &
  \multicolumn{1}{c|}{8.4} &
  47.0 \\
  Granite 34B &
   &
  \multicolumn{1}{c|}{52.1} &
  \multicolumn{1}{c|}{68.0} &
  \multicolumn{1}{c|}{26.4} &
  \multicolumn{1}{c|}{61.0} &
  \multicolumn{1}{c|}{8.6} &
  \textbf{58.0}\\
Codellama 34B &
   &
  \multicolumn{1}{c|}{48.0} &
  \multicolumn{1}{c|}{59.0} &
  \multicolumn{1}{c|}{27.9} &
  \multicolumn{1}{c|}{53.0} &
  \multicolumn{1}{c|}{9.1} &
  \textbf{58.0} \\ 
\textbf{\faTrophy{} Llama3 70B} &
   &
  \multicolumn{1}{c|}{\textbf{56.0}} &
  \multicolumn{1}{c|}{\textbf{71.0}} &
  \multicolumn{1}{c|}{\textbf{32.4}} &
  \multicolumn{1}{c|}{\textbf{63.2}} &
  \multicolumn{1}{c|}{\textbf{14.6}} &
  {56.9} \\\midrule
Llama3 8B &
  \multirow{5}{*}{Python} &
  \multicolumn{1}{c|}{13.7} &
  \multicolumn{1}{c|}{43.0} &
  \multicolumn{1}{c|}{10.2} &
  \multicolumn{1}{c|}{42.0} &
  \multicolumn{1}{c|}{\textbf{11.6}} &
  43.0 \\
Granite 8B &
   &
  \multicolumn{1}{c|}{10.2} &
  \multicolumn{1}{c|}{54.0} &
  \multicolumn{1}{c|}{11.9} &
  \multicolumn{1}{c|}{58.0} &
  \multicolumn{1}{c|}{11.1} &
  \textbf{55.0} \\
Granite 20B &
   &
  \multicolumn{1}{c|}{14.6} &
  \multicolumn{1}{c|}{45.0} &
  \multicolumn{1}{c|}{11.7} &
  \multicolumn{1}{c|}{67.0} &
  \multicolumn{1}{c|}{7.3} &
  44.0 \\
  Granite 34B &
   &
  \multicolumn{1}{c|}{17.7} &
  \multicolumn{1}{c|}{54.0} &
  \multicolumn{1}{c|}{13.9} &
  \multicolumn{1}{c|}{67.0} &
  \multicolumn{1}{c|}{10.6} &
  46.0 \\
Codellama 34B &
   &
  \multicolumn{1}{c|}{13.7} &
  \multicolumn{1}{c|}{49.0} &
  \multicolumn{1}{c|}{11.6} &
  \multicolumn{1}{c|}{53.0} &
  \multicolumn{1}{c|}{8.4} &
  44.0 \\ 
\textbf{\faTrophy{} Llama3 70B} &
   &
  \multicolumn{1}{c|}{\textbf{24.7}} &
  \multicolumn{1}{c|}{\textbf{57.2}} &
  \multicolumn{1}{c|}{\textbf{18.9}} &
  \multicolumn{1}{c|}{\textbf{70.4}} &
  \multicolumn{1}{c|}{\textbf{14.9}} &
  52.1 \\\midrule
Llama3 8B &
  \multirow{5}{*}{NL} &
  \multicolumn{1}{c|}{30.2} &
  \multicolumn{1}{c|}{63.0} &
  \multicolumn{1}{c|}{24.5} &
  \multicolumn{1}{c|}{56.0} &
  \multicolumn{1}{c|}{9.6} &
  57.0 \\
Granite 8B &
   &
  \multicolumn{1}{c|}{52.3} &
  \multicolumn{1}{c|}{59.0} &
  \multicolumn{1}{c|}{42.1} &
  \multicolumn{1}{c|}{61.0} &
  \multicolumn{1}{c|}{11.1} &
  58.0 \\
  Granite 20B &
   &
  \multicolumn{1}{c|}{65.4} &
  \multicolumn{1}{c|}{54.0} &
  \multicolumn{1}{c|}{46.0} &
  \multicolumn{1}{c|}{48.0} &
  \multicolumn{1}{c|}{10.9} &
  \textbf{60.0} \\
Granite 34B &
   &
  \multicolumn{1}{c|}{69.7} &
  \multicolumn{1}{c|}{55.0} &
  \multicolumn{1}{c|}{55.1} &
  \multicolumn{1}{c|}{46.0} &
  \multicolumn{1}{c|}{10.9} &
  56.0 \\
Codellama 34B &
   &
  \multicolumn{1}{c|}{60.4} &
  \multicolumn{1}{c|}{57.0} &
  \multicolumn{1}{c|}{40.6} &
  \multicolumn{1}{c|}{57.0} &
  \multicolumn{1}{c|}{8.69} &
  50.0 \\ 
\textbf{\faTrophy{} Llama3 70B} &
   &
  \multicolumn{1}{c|}{\textbf{75.2}} &
  \multicolumn{1}{c|}{\textbf{67.7}} &
  \multicolumn{1}{c|}{\textbf{57.2}} &
  \multicolumn{1}{c|}{\textbf{64.2}} &
  \multicolumn{1}{c|}{\textbf{13.4}} &
  58.1 \\\bottomrule
\end{tabularx}
\caption{Zero shot results for both the experiments. Models scoring the highest accuracy the majority of times across all output representations for a particular schema are labeled with \faTrophy{}. Gen Acc represents the accuracy of valid samples for DSL generation experiment. Val Acc represents the accuracy of the binary classification validation experiment.}

\label{table:zeroshotmainresults}
\end{table*}

\paragraph{Description:} \label{exp1desc} Given the schema, the experiment (see Listing \ref{lst:task1}) aims to produce a compliant data sample in DSL code format. 
We draw inspiration from several use cases (see Appendix \ref{sec:taskmotivation}), including synthesizing schema-compliant data from LLMs' parametric memory to train and evaluate smaller-sized models \cite{song2020lightpafftwostagedistillationframework} and generating diverse sets of samples to be used in product test pipelines. For reliable DSL code generation, LLMs need to be schema-aware.

\paragraph{Dataset:}\label{codegen-dataset}  We synthetically prepare $602$ schemas for each of the $5$ representations having combinations of various constraints (Appendix \ref{sec:schemaex}). First, we prepare JSON schemas using our combinatorial tool to generate a good mix of constraints. A combinatorial data generation tool factors in constraints of interest, constraint-specific information, and combinatorial preferences to generate the schemas. We then convert each JSON schema to XML and YAML schemas using openly available automatic lossless language-to-language translation tools. Further, we include resource-rich general-purpose language - Python using the Pydantic library generated using the Gemini-1.0-pro \citep{team2023gemini} model as a code translation task. We extend our evaluation to NL representation generated using rule-based templates. We\footnote{The schemas are manually validated by the paper's authors.} ensure equivalence of the generated schemas across languages. We plan to open-source all the scripts used for data preparation. Table \ref{table:schema_length} gives details regarding schema token length.


\paragraph{Evaluation metric:} Each schema-compliant code output LLM generates is awarded one point where schema compliance is checked using a schema validator tool. We then utilize the accuracy metric (Gen Acc) over all samples to benchmark performance across the models. Additionally, we also report the percentage of samples generated with the invalid root data type (RTV\%) and invalid samples (IS\%) in Table \ref{table:task1isandrtvappendix}. The root data type is the data type of the whole DSL sample. For example, the root data type of sample represented in Listing \ref{lst:task1} is \emph{array}. For IS and RTV metrics, the lesser the number, the better the performance.

\paragraph{Experimental setup:}\label{expriment} We report greedy decoding results since it performed slightly better than beam search with a beam width of 3. We perform inference for all the models in $bfloat16$ precision and a max new token limit of 1024 tokens.


\paragraph{Prompts:} We experiment with zero- and 3-shot prompting for each model. For 3-shot prompting, we identify errors from the zero-shot setting, then select shots similar to the most frequent errors. We observe that most errors made by all the models are regarding short schema and the schema having root type of array as shown in sample \ref{lst:task1}. An example of a 3-shot prompt for a DSL generation experiment is shown below. Examples of prompts are in Appendix \ref{lst:task1}.

\begin{table*}[!ht]
\setlength\tabcolsep{14pt}
\scriptsize
\centering
\begin{tabularx}{\textwidth}{c|c|cccccc}
\toprule
 &
   &
  \multicolumn{6}{c}{Output Representation} \\ \midrule
 &
   &
  \multicolumn{2}{c|}{JSON} &
  \multicolumn{2}{c|}{YAML} &
  \multicolumn{2}{c}{XML} \\ \midrule
Model &
  Schema &
  \multicolumn{1}{c|}{Gen Acc} &
  \multicolumn{1}{c|}{Val Acc} &
  \multicolumn{1}{c|}{Gen Acc} &
  \multicolumn{1}{c|}{Val Acc} &
  \multicolumn{1}{c|}{Gen Acc} &
    Val Acc \\ \midrule
Llama3 8B &
  \multirow{5}{*}{JSON} &
  \multicolumn{1}{c|}{48.3} &
  \multicolumn{1}{c|}{71.2} &
  \multicolumn{1}{c|}{46.6} &
  \multicolumn{1}{c|}{68.1} &
  \multicolumn{1}{c|}{39.2} &
  64.1 \\
Granite 8B &
   &
  \multicolumn{1}{c|}{51.2} &
  \multicolumn{1}{c|}{69.2} &
  \multicolumn{1}{c|}{52.3} &
  \multicolumn{1}{c|}{66.1} &
  \multicolumn{1}{c|}{47.8} &
  65.8 \\
Granite 20B &
   &
  \multicolumn{1}{c|}{58.3} &
  \multicolumn{1}{c|}{73.5} &
  \multicolumn{1}{c|}{56.4} &
  \multicolumn{1}{c|}{72.3} &
  \multicolumn{1}{c|}{50.2} &
  68.2 \\
Granite 34B &
   &
  \multicolumn{1}{c|}{66.3} &
  \multicolumn{1}{c|}{76.2} &
  \multicolumn{1}{c|}{64.5} &
  \multicolumn{1}{c|}{75.4} &
  \multicolumn{1}{c|}{51.3} &
  73.2 \\
Codellama 34B &
   &
  \multicolumn{1}{c|}{65.1} &
  \multicolumn{1}{c|}{75.1} &
  \multicolumn{1}{c|}{63.4} &
  \multicolumn{1}{c|}{73.2} &
  \multicolumn{1}{c|}{50.6} &
  71.2 \\
\textbf{\faTrophy{} Llama3 70B} &
   &
  \multicolumn{1}{c|}{\textbf{70.1}} &
  \multicolumn{1}{c|}{\textbf{79.3}} &
  \multicolumn{1}{c|}{\textbf{69.4}} &
  \multicolumn{1}{c|}{\textbf{77.9}} &
  \multicolumn{1}{c|}{\textbf{58.6}} &
  \textbf{74.2}\\ \midrule
Llama3 8B &
  \multirow{5}{*}{XML} &
  \multicolumn{1}{c|}{46.6} &
  \multicolumn{1}{c|}{65.8} &
  \multicolumn{1}{c|}{42.3} &
  \multicolumn{1}{c|}{63.4} &
  \multicolumn{1}{c|}{36.6} &
  60.1 \\
Granite 8B &
   &
  \multicolumn{1}{c|}{46.2} &
  \multicolumn{1}{c|}{64.8} &
  \multicolumn{1}{c|}{44.5} &
  \multicolumn{1}{c|}{63.2} &
  \multicolumn{1}{c|}{34.5} &
  57.3 \\
Granite 20B &
   &
  \multicolumn{1}{c|}{50.4} &
  \multicolumn{1}{c|}{66.7} &
  \multicolumn{1}{c|}{48.2} &
  \multicolumn{1}{c|}{64.1} &
  \multicolumn{1}{c|}{36.4} &
  56.1 \\
Granite 34B &
   &
  \multicolumn{1}{c|}{52.3} &
  \multicolumn{1}{c|}{68.5} &
  \multicolumn{1}{c|}{51.1} &
  \multicolumn{1}{c|}{63.4} &
  \multicolumn{1}{c|}{39.2} &
  53.2 \\
Codellama 34B &
   &
  \multicolumn{1}{c|}{49.2} &
  \multicolumn{1}{c|}{66.2} &
  \multicolumn{1}{c|}{49.2} &
  \multicolumn{1}{c|}{63.2} &
  \multicolumn{1}{c|}{35.1} &
  52.1 \\ 
\textbf{\faTrophy{} Llama3 70B} &
   &
  \multicolumn{1}{c|}{\textbf{56.4}} &
  \multicolumn{1}{c|}{\textbf{70.3}} &
  \multicolumn{1}{c|}{\textbf{55.6}} &
  \multicolumn{1}{c|}{\textbf{68.2}} &
  \multicolumn{1}{c|}{\textbf{43.6}} &
  \textbf{66.3} \\ \midrule
Llama3 8B &
  \multirow{5}{*}{YAML} &
  \multicolumn{1}{c|}{46.7} &
  \multicolumn{1}{c|}{67.2} &
  \multicolumn{1}{c|}{45.3} &
  \multicolumn{1}{c|}{64.2} &
  \multicolumn{1}{c|}{43.5} &
  63.2 \\
Granite 8B &
   &
  \multicolumn{1}{c|}{48.1} &
  \multicolumn{1}{c|}{65.2} &
  \multicolumn{1}{c|}{46.2} &
  \multicolumn{1}{c|}{61.2} &
  \multicolumn{1}{c|}{44.2} &
  61.2 \\
Granite 20B &
   &
  \multicolumn{1}{c|}{52.3} &
  \multicolumn{1}{c|}{68.9} &
  \multicolumn{1}{c|}{49.7} &
  \multicolumn{1}{c|}{66.7} &
  \multicolumn{1}{c|}{47.8} &
  65.1 \\
Granite 34B &
   &
  \multicolumn{1}{c|}{54.2} &
  \multicolumn{1}{c|}{67.7} &
  \multicolumn{1}{c|}{51.3} &
  \multicolumn{1}{c|}{65.3} &
  \multicolumn{1}{c|}{45.3} &
  56.4\\
Codellama 34B &
   &
  \multicolumn{1}{c|}{56.8} &
  \multicolumn{1}{c|}{66.4} &
  \multicolumn{1}{c|}{50.2} &
  \multicolumn{1}{c|}{64.3} &
  \multicolumn{1}{c|}{47.8} &
  56.2 \\ 
\textbf{\faTrophy{} Llama3 70B} &
   &
  \multicolumn{1}{c|}{\textbf{60.4}} &
  \multicolumn{1}{c|}{\textbf{76.3}} &
  \multicolumn{1}{c|}{\textbf{57.3}} &
  \multicolumn{1}{c|}{\textbf{69.1}} &
  \multicolumn{1}{c|}{\textbf{49.6}} &
  \textbf{68.3} \\ \midrule
Llama3 8B &
  \multirow{5}{*}{Python} &
  \multicolumn{1}{c|}{43.2} &
  \multicolumn{1}{c|}{60.1} &
  \multicolumn{1}{c|}{41.1} &
  \multicolumn{1}{c|}{58.9} &
  \multicolumn{1}{c|}{39.2} &
  57.6 \\
Granite 8B &
   &
  \multicolumn{1}{c|}{45.1} &
  \multicolumn{1}{c|}{60.5} &
  \multicolumn{1}{c|}{46.7} &
  \multicolumn{1}{c|}{59.4} &
  \multicolumn{1}{c|}{37.4} &
  56.0 \\
Granite 20B &
   &
  \multicolumn{1}{c|}{48.2} &
  \multicolumn{1}{c|}{57.2} &
  \multicolumn{1}{c|}{45.9} &
  \multicolumn{1}{c|}{57.8} &
  \multicolumn{1}{c|}{38.4} &
   58.2 \\
Granite 34B &
   &
  \multicolumn{1}{c|}{50.6} &
  \multicolumn{1}{c|}{59.2} &
  \multicolumn{1}{c|}{47.1} &
  \multicolumn{1}{c|}{55.6} &
  \multicolumn{1}{c|}{41.3} &
  57.3 \\
Codellama 34B &
   &
  \multicolumn{1}{c|}{47.2} &
  \multicolumn{1}{c|}{56.4} &
  \multicolumn{1}{c|}{45.3} &
  \multicolumn{1}{c|}{57.2} &
  \multicolumn{1}{c|}{39.2} &
  55.1 \\ 
\textbf{\faTrophy{} Llama3 70B} &
   &
  \multicolumn{1}{c|}{\textbf{56.2}} &
  \multicolumn{1}{c|}{\textbf{65.1}} &
  \multicolumn{1}{c|}{\textbf{50.7}} &
  \multicolumn{1}{c|}{\textbf{64.2}} &
  \multicolumn{1}{c|}{\textbf{43.4}} &
  \textbf{60.6} \\ \bottomrule
\end{tabularx}
\caption{Few shot results for generation (3 shots) and validation (2 shots) experiments. Models scoring the highest accuracy the majority number of times across all output representations for a particular schema are labeled with \faTrophy{}. Gen Acc represents the accuracy of valid samples for DSL generation experiment. Val Acc represents the accuracy of the binary classification validation experiment.}

\label{table:few_shotappendix}
\end{table*}

\subsection{DSL Validation}\label{task2para}

\paragraph{Description:} \label{exp2desc} There is a growing body of work \cite{hada-etal-2024-large} on showing promising usage of LLMs as evaluators in many tasks. On similar lines, given the DSL sample and schema to validate, this experiment (see Listing \ref{lst:task2}) aims to determine the validity of the provided sample against the constraints through boolean question answering (QA). Also, the experiment is highly motivated from various use cases (see Appendix \ref{sec:taskmotivation}) and throws light on LM's understanding of the relation between requirements and output in various representations. 


\paragraph{Dataset:} \label{codegen-dataset}  We synthetically prepare $602$ schemas across $5$ representations having combinations of hard and soft constraints. First, we prepare JSON schemas using our combinatorial tool to generate a good mix of constraints. We then convert each JSON schema to XML and YAML schemas using automated tools to ensure equivalence across representations. Further, we include Python representation using the Pydantic library as a resource-rich general-purpose language in our evaluation generated using the Gemini-1.0-pro \citep{team2023gemini} model as a code translation task. We extend our evaluation to natural language representation generated using rule-based templates over the JSON schema. We\footnote{The generated Python samples are manually validated by the paper's authors.} ensure equivalence of the generated schemas across languages by manually eyeballing the samples.


\begin{lstlisting}[language=json, caption=In the JSON sample{,} values for fields \emph{stingo} and \emph{anisic} do not adhere to schema constraints. But the Granite 34B model gives the incorrect answer (highlighted in yellow) as \emph{yes}., label={lst:task2}, escapechar=@]
Question:
Does the JSON sample { "tamil": false, "baser": null, "anisic": 1906.34, "stingo": "officiis tellus. illum modi odit quas mattis nunc", "pigheadedness": 52.0 } adhere to all the constraints defined in JSON format schema
{ 
  "type": "object",
  "properties": {
    "tamil": { "type": "boolean" },
    "baser": { "type": "null" },
    "anisic": { "type": "number", "multipleOf": 17.02 },
    "stingo": { "type": "string", "maxLength": 20 },
    "pigheadedness": {"type": "number", "exclusiveMinimum": 27.65410407394338, "maximum": 93.85523810367313 } },
  "additionalProperties": false
}
Respond to yes or no.
Answer:
```
@\highlight{yes}@
@\highlight{```}@
\end{lstlisting}

\paragraph{Evaluation metric:} Since it is a boolean QA experiment, we use Macro average F1 (see Table \ref{table:task2macrof1appendix}) and Accuracy (Val Acc) as evaluation metrics (see Table \ref{table:zeroshotmainresults}).

\paragraph{Experimental setup:} The decoding strategy used here is similar to the data generation experiment as mentioned in Section \ref{expriment}. We perform inference in $bfloat16$ precision and a max new token limit of 1024 tokens. For beam search decoding, we use the beam width of 3.

\paragraph{Prompts:} The goal of this experiment is to answer \emph{yes} or \emph{no}. We experiment with zero- and few-shot prompting. With few shot prompting, we provide one example each of \emph{yes} and \emph{no} answers. Results for few-shot prompting and examples of prompts are given in Appendix (Table \ref{table:few_shotappendix}).

\section{Format Efficacy}
\subsection{Objective}
To identify the most effective schema representation and output format for system-level programming while employing language models. Since schemas can be represented in various structured formats, including JSON, YAML, XML, Python, and even NL, determining which format best enables constraint adherence for language models while balancing context-length costs is critical.
\subsection{RQ1: Which format is optimally suited for constraint and output representation?}\label{formatrq1}

\paragraph{Finding 1.}In the data as code generation experiment (section \ref{task1para}), models best understand (Table \ref{table:zeroshotmainresults}) NL across all outputs. At the same time, JSON and YAML schemas perform well (Table \ref{table:few_shotappendix}) for constraints in code despite their limited presence in pre-training data. Surprisingly, models struggle with constraints in Python, likely due to a bias toward generating general-purpose Python code rather than schema-specific patterns. In contrast, JSON and YAML schemas benefit from their rigid structures and alignment with schema-centric applications, making them easier for models to interpret. 

\paragraph{Finding 2.} Using the same schema and output representation does not always enhance performance. For instance, in Table \ref{table:few_shotappendix}, YAML as schema and JSON as output representation performed better than YAML for both representations. 

\paragraph{Finding 3.} Although NL representation excels in generation experiments, it degrades the validation performance of larger models like 70B. Like generation experiment, models perform sub-optimally when schema and output representations are the same. In line with the first experiment, XML stands as a challenging language for models. The Llama3 70B model performs best in validation as in the first experiment, with other models hovering around 50\% Val Acc, likely reflecting the random choice given the binary nature of the experiment. Smaller models, particularly the Llama3-8B with natural language representation, show notable improvement, as its pre-training combines NL and code.

\paragraph{Key takeaway.} NL is a favorable language for schema representation, however, since its possible that enterprises lean more toward structured languages for better interoperability in which case JSON and YAML are ideal candidates for schema representation with JSON being favourable candidate for output representation. Nonetheless, the inconsistency in performance across experiments and model sizes underscores need for better schema comprehension and improved training strategies for NLP tasks involving validation.



\subsection{RQ2: What is the trade-off between performance and context length cost?}

\paragraph{Findings.} From section \ref{formatrq1} key takeaway, JSON and YAML are ideal candidates for schema representation which form the context to the LLM. From Table \ref{table:schema_length}, representing schema in YAML on an average takes $\sim$35\% less tokens than JSON. However, while choosing YAML would mean taking a drop of $\sim$14\% in Gen Acc and $\sim$4\% in Val Acc performance compared to JSON.

\paragraph{Key takeaway.} Enterprises should be cognizant of such tradeoff and choose ideal representation that fits their use case. Further, better tokenizer training techniques might lead to lower token expenditure for the desired representation.


\section{Constraint Efficacy}
\subsection{Objective}
To examine how language models handle various types of constraints embedded within schemas. Enterprise schemas enforce structural (e.g., required fields, data types) and semantic (e.g., dependencies, value constraints) rules.
\subsection{RQ1: How does performance vary across different types of constraints?}

\paragraph{Findings.} The analysis of the results shows that LLaMA3 8B and 70B exhibit similar patterns of missing constraints when generating JSON samples from a given schema (Table \ref{table:errortypeanalysis}). In particular, constraints such as type, multiple, and exclusiveMinimum are often missing, while constraints such as maximum, additionalProperties, and minimum are more frequently followed. The high error rate in fundamental constraints \emph{type} can be because training data contains many JSON-like samples where \emph{type} is implicit rather than explicitly stated. The reason behind missing constraints like \emph{exclusiveMinimum} and \emph{multipleOf} may be because they involve high numerical precision. LLMs treat numbers as tokens, leading to potential rounding errors or incorrect enforcement.

\paragraph{Key takeaway.} LLMs struggle with numerical constraints underscoring need for better techniques throughout the stack from tokenizer to training. For enterprises, a rudimentary solution is to integrate constrained decoding or use post-processing validation to correct missing constraints after generation.

\subsection{RQ2: What are the ideal positions for constraints in the schema for better adherence?}

\paragraph{Findings.} We categorize the constraints of the schema into three sections based on tokens: beginning (first 30\%), middle (next 40\%), and end (last 30\%). Later, we perform a needle-in-the-haystack experiment for the data-as-code generation. The heatmap in Figure \ref{needle figure} shows the statistics of constraints missed at every position for JSON to JSON generation. It reveals a consistent trend where models struggle the most with constraints positioned at the beginning of the schema, followed by the middle. In contrast, constraints at the end are least frequently missed. This suggests that models may prioritize constraints appearing later in the schema, likely due to the left-to-right decoding nature of autoregressive models, causing early constraints to be overwritten or ignored. We also observe that constraints in the middle position of the schema are frequently missed. This aligns with previous findings that the middle part of the long context is often missed \cite{liu2024lost}. For the data validation task, we analyze attention maps, which reveal a similar trend where the model pays less attention to the middle part of the schema (Figure \ref{attention_maps}). 

\paragraph{Key takeaway.}This suggests that important constraints should be placed at the end of the schema or the beginning for longer schemas, depending on the use case.

\section{Related Work}

\paragraph{Generation:} There is extensive work \cite{DBLP:conf/iclr/MuennighoffLZZH24,cassano:multipl-e} on evaluating capabilities of LLMs for various code tasks such as code completion, translation, etc, for resource-rich languages like Python. 
Despite there being work \cite{cassano:multipl-e} on multi-lingual code, there is scant attention to low-resource languages such as DSLs, though having crucial importance. One notable work \cite{he2024doespromptformattingimpact}, studies the bearing of prompt format in DSLs with LLM performance, however, does not include impact of output formats and controllability aspect in terms of code constraints crucial for enterprises. Further, using LLMs as evaluators for low-resource languages is gaining interest, however limited, mainly focusing on languages like XML and INI \cite{DBLP:journals/corr/abs-2310-09690}.



\paragraph{Controllability of LLMs:} While LLMs can handle coarse-grained constraints like sentiment, they struggle with fine-grained constraints, such as ending a text with a specific word \cite{sun-etal-2023-evaluating}. Code schemas often require such fine-grained control, and to our knowledge, we are the first to explore LLM controllability for constraints in code.

\section{Conclusion}

We evaluate LLMs for system-level programming across two key dimensions: Format Efficacy and Constraint Efficacy. Format efficacy examines how LLMs handle different constraint formats, while constraint efficacy assesses their performance on various schema constraints within a format. We conduct two novel experiments to study these aspects: Data as Code generation and DSL validation. We evaluate LLMs across $5$ schema(YAML, JSON, Python, XML, NL) and $3$ output formats(YAML, JSON, XML). Our findings reveal that model performance does not directly correlate with a language's presence in pre-training data. JSON and YAML are best suited for system-level programming, and enterprises should convert Python and XML formats to one of these for better LLM performance. We also observe that schema constraint locality affects performance, with constraints in the start and middle being most frequently violated. Placing critical constraints at the end improves reliability. We hope our work drives innovation in improving LLM capabilities for crucial industry use case of system-level programming involving code constraints.



\section{Limitations}
While we explore the DSL validation task by generating \emph{yes} or \emph{no}, exploring the model's reasoning can give a more comprehensive analysis of LLM's understanding. Further, one can include more complex constraints in the future for general-purpose programming languages, like coding style constraints to write code along with natural language prompts and schema.

\section*{Ethics Statement}
Custom-created datasets have been created synthetically using open-source tools. The language models, tools, and frameworks used for evaluation are open source and can be used without copyright issues.


\bibliography{custom}

\appendix

\section{Appendix}
\label{sec:appendix}

\subsection{Prompts}\label{prompts}

This section defines the prompts which are used for models. We report different prompts for every model tried here and report the best-performing prompt results.
Generally, the model consists of a System Prompt followed by a prompt template specific to the model.

\subsubsection{Common prompt}
For zero shot inference, we use a common prompt as it is for all the models irrespective of the model's prompt format and we observe best results for Task-1 with this prompt. The prompt is as follows.\\

\begin{lstlisting}[language=json, caption=common prompt, label={lst:common}, escapechar=@]
Write an {input_representation} sample with field values as per the {output_representation} format schema given below.

{schema}

{output_representation} sample:
```
\end{lstlisting}

\subsubsection{Granite model family}

The granite model generally follows the question-answering format. Task-1 prompts for granite family models are as follows.\\

\noindent\textbf{System prompt:}\\
System:\\
You are an intelligent AI programming assistant, utilizing a Granite code language model developed by IBM. Your primary function is to assist users in code explanation, code generation and other software engineering tasks. You MUST follow these guidelines: - Your responses must be factual. Do not assume the answer is \emph{yes} when you do not know, and DO NOT SHARE FALSE INFORMATION. - You should give concise answers. You should follow the instruction and provide the answer in the specified format and DO NOT SHARE FALSE INFORMATION.

\noindent\textbf{Prompt 2:}\\
\begin{lstlisting}[language=json, caption=QA-prompt-1, label={lst:QA-1}, escapechar=@]
{System prompt}

Question:
Write an {input_representation} sample with field values as per the {input_representation} format schema given below.

{schema}

Answer:
```
\end{lstlisting}

\noindent\textbf{Prompt 3:}\\
\begin{lstlisting}[language=json, caption=QA-prompt-2, label={lst:QA-2}, escapechar=@]
{System prompt}

Question:
Write an {input_representation} sample with field values as per the {output_representation} format schema given below. Please wrap your code
answer using ```

{schema}

Answer:
```
\end{lstlisting}

\noindent\{output\_representation\} and \{input\_representation\} are the variables where \{input\_representation\} take the values JSON, YAML, XML, Python, and natural language. \{output\_representation\} takes the values JSON, YAML, and XML.

\subsubsection{Llama family}

For codellama $34B$ model we wrap the common prompt in [INST] and [/INST] tags. For the llama3-8B model, we use the System prompt along with user tags \footnote{\url{https://huggingface.co/meta-llama/Meta-Llama-3-8B-Instruct}}.

\noindent\textbf{System prompt:}
You are a helpful, respectful, and honest assistant. Always answer as helpfully as possible, while being safe. Your answers should not include any harmful, unethical, racist, sexist, toxic, dangerous, or illegal content. 
Please ensure that your responses are socially unbiased and positive.
If a question does not make any sense or is not factually coherent, explain why instead of answering something not correct. If you don't know the answer to a question, please don't share false information.

\noindent Other than this, similar to the granite family, we try Question answering format and instruction to wrap the output in quotes (```).


\noindent\textbf{Few shot prompt}\\\\
\begin{lstlisting}[language=json, caption=Few shot prompt, label={lst:few-shot}, escapechar=@]
{System prompt}

Your task is to write a JSON sample with field values as per JSON format schema.
You are given a few examples demonstrating the same.

JSON format schema:
{
    "type": "array",
    "contains": {
        "type": "boolean"
    },
    "minContains": 0
}
JSON sample:
```
[true, true, false]
```
JSON format schema:
{
    "type": "string",
    "format": "idn-email"
}
JSON sample:
```
"hchavez@example.org"
```
JSON format schema:
{
    "type": "array",
    "items": {
        "type": "number",
        "multipleOf": 5.82,
        "exclusiveMinimum": 3.069158195370172
    }
}

JSON sample:
```
\end{lstlisting}

\begin{figure*}
  \includegraphics[width=\textwidth,height=15cm]{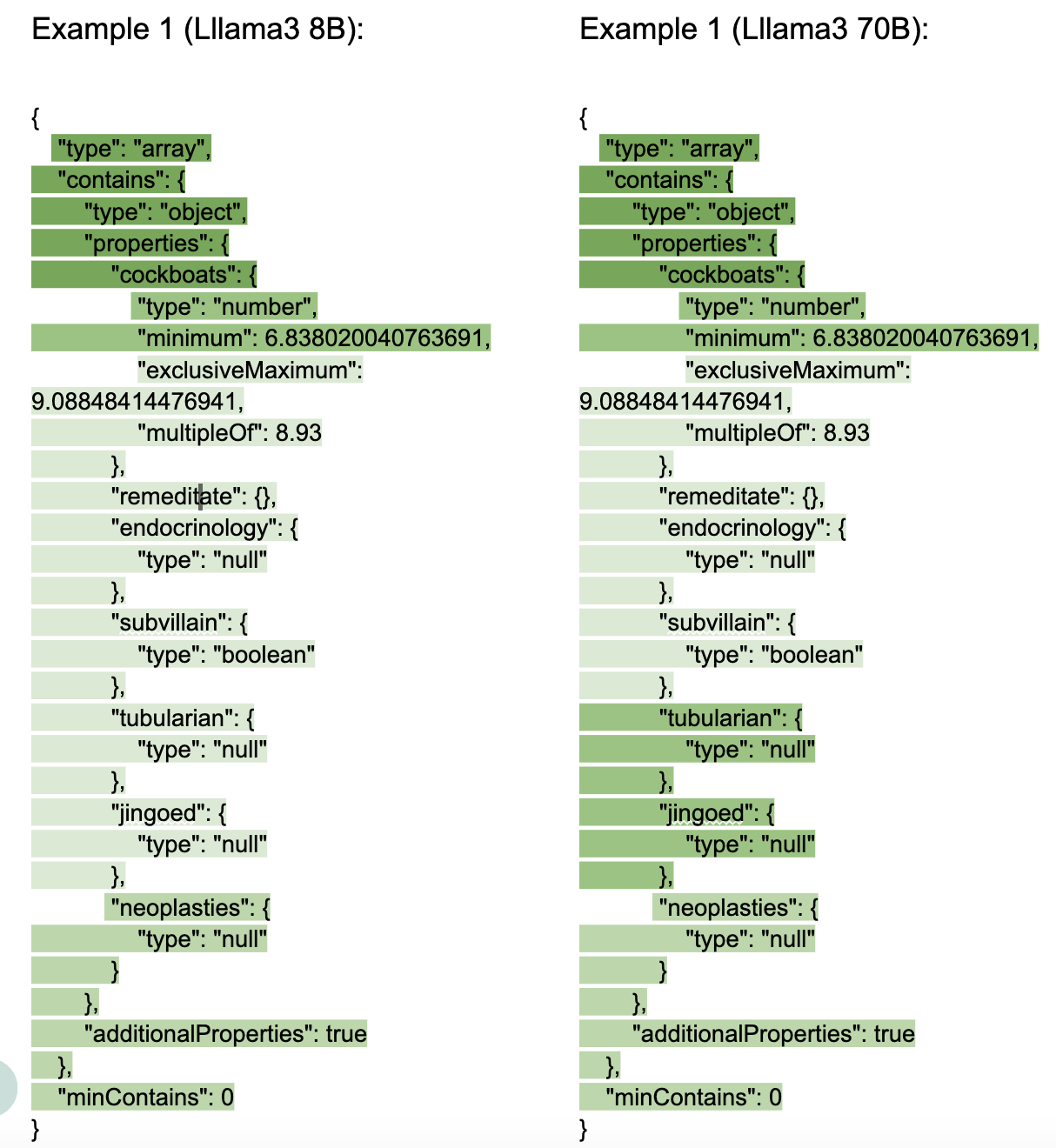}
  \caption{Attention maps for LLama3 8B and 70B model for Data validation experiment. The more the intensity of color, the more attention is given to that part of input by the model.}
\label{attention_maps}
\end{figure*}

\subsection{Limitations of Constrained Decoding} \label{sec:cd}
This section outlines some common problems with constrained decoding and emphasizes why it cannot be a complete and viable solution for factoring in schemas to generate compliant text using language models.

\subsubsection{Inference Performance Bottleneck}
Constrained decoding often negatively affects inference throughput, widely mentioned as one of the major drawbacks in many works \cite{10.5555/3666122.3668959, pimparkhede-etal-2024-doccgen, geng-etal-2023-grammar} due to involvement of token-level operations keeping track of the schema constraints and tokens generated so far. This latency can be a factor of the complexity of the schema, tokens generated so far, and the nature of the constrained decoding implementation. Further, advances such as batched inference \footnote{\url{https://github.com/microsoft/batch-inference}} are not yet there for constrained decoding limiting their scalability and practical use.

\subsubsection{Complex Engineering Effort}
Implementing a constrained decoding system can involve instrumenting at the decoding phase of the language model while keeping track of the tokens generated so far and structured schema adherence which can involve implementation specific to a schema representation and may not be possible to generalize to any schema representation. For instance, most of the openly available constrained decoding systems \footnote{\url{https://github.com/outlines-dev/outlines}} have limited support and not generalized to various schemas such as XML and output formats such as YAML and others. It is worthwhile to note that some approaches tend to convert scehmas to context free grammars, however, this approach is possible with common schema representations such as Python pydantic. Additionally, implementing such a system requires deep domain expertise.

\begin{table*}[h!]
\setlength\tabcolsep{12pt}
\small
\centering
\begin{tabularx}{\textwidth}{X|X|X}
\toprule
\textbf{Constraint} & \textbf{Llama3 8B} & \textbf{Llama3 70B} \\
\midrule
type & 302 &	49 \\
\midrule
exclusiveMinimum	& 18	& 44 \\
\midrule
multipleOf	&170&	42 \\
\midrule
minLength	&47&	21 \\
\midrule
contains	&22&	12 \\
\midrule
exclusiveMaximum	&22&	12 \\
\midrule
maximum	&11&	2 \\
\midrule
maxLength	&7&	19 \\
\midrule
additionalProperties	&4&	0 \\
\midrule
minimum	&4&	15 \\

\bottomrule
\end{tabularx}
\caption{Both the models, least and best performing, irrespective of their performance, show a similar distribution of mistakes for each constraint.}
\label{table:errortypeanalysis}
\end{table*}

\subsubsection{Model Performance Bottleneck}
LLMs have multiple failure modes that can likely be triggered through constrained decoding. Many works show that LLMs are sensitive to the text being fed into them and often deteriorate the model's performance. Some examples being the reverse curse from \cite{DBLP:conf/iclr/BerglundTKBSKE24}, where LLM understanding "A is B" may not guarantee to learn "B is A". Another work \cite{10.5555/3692070.3692325} shows that the order of the premises can have a substantial impact on the performance often affecting negatively. Such failures can be triggered when the natural flow of text generation is interrupted through constrained decoding over autoregressive generation. The problem can worsen when it involves mixed generation of structured output and unstructured NL text.

\subsubsection{Limited Scope}
Since constrained decoding needs access to the decoding phase of the language model, its often not possible to apply such decoding to hosted or gated LLM deployments. 

Applying constrained decoding to some common use cases is not obvious. Given $n$ structured schemas from $s_1$ to $s_n$, unstructured NL text output as $k$ and structured output as $u$. Common use cases in natural language processing (NLP) such as summarization involve the following input-output relationship. For some arbitrary schema $i$, $s_i$ $\rightarrow$ $u$. Further typical use cases involve factoring in $n$ multiple schemas and generate $m$ multiple structured outputs $(s_1...s_n)$  $\rightarrow$ $(k_1...k_m)$.

Employing constrained decoding in such use cases is not viable since in the first use case, tasks that output $u$ cannot leverage constrained decoding and schema has to go into LLMs as input. When multiple schemas and structured outputs are involved, its not obvious to choose the right schema for decoding a particular structured output. Such common use cases substantially limit the scope of using constrained decoding.

\begin{table*}[!ht]
\setlength\tabcolsep{14pt}
\scriptsize
\begin{tabularx}{\textwidth}{c|c|cccccc}
\toprule
 &
   &
  \multicolumn{6}{c}{Output Representation} \\ \midrule
 &
   &
  \multicolumn{2}{c|}{JSON} &
  \multicolumn{2}{c|}{YAML} &
  \multicolumn{2}{c}{XML} \\ \midrule
Model &
  Schema &
  \multicolumn{1}{c|}{IS (\%)} &
  \multicolumn{1}{c|}{RTV (\%)} &
  \multicolumn{1}{c|}{IS (\%)} &
  \multicolumn{1}{c|}{RTV (\%)} &
  \multicolumn{1}{c|}{IS (\%)} &
  RTV (\%) \\ \midrule
Llama3 8B &
  \multirow{5}{*}{JSON} &
  \multicolumn{1}{c|}{\textbf{1.9}} &
  \multicolumn{1}{c|}{50.1} &
  \multicolumn{1}{c|}{1.8} &
  \multicolumn{1}{c|}{49.8} &
  \multicolumn{1}{c|}{\textbf{1.6}} &
  73.9 \\
Granite 8B &
   &
  \multicolumn{1}{c|}{2.9} &
  \multicolumn{1}{c|}{31.0} &
  \multicolumn{1}{c|}{2.8} &
  \multicolumn{1}{c|}{57.3} &
  \multicolumn{1}{c|}{17.1} &
  \textbf{70.26} \\
Granite 20B &
   &
  \multicolumn{1}{c|}{13.9} &
  \multicolumn{1}{c|}{\textbf{15.6}} &
  \multicolumn{1}{c|}{2.3} &
  \multicolumn{1}{c|}{\textbf{38.0}} &
  \multicolumn{1}{c|}{7.9} &
  71.92 \\
Granite 34B &
   &
  \multicolumn{1}{c|}{2.6} &
  \multicolumn{1}{c|}{23.5} &
  \multicolumn{1}{c|}{2.6} &
  \multicolumn{1}{c|}{48.6} &
  \multicolumn{1}{c|}{4.1} &
  73.08 \\
Codellama 34B &
   &
  \multicolumn{1}{c|}{3.6} &
  \multicolumn{1}{c|}{17.9} &
  \multicolumn{1}{c|}{\textbf{1.8}} &
  \multicolumn{1}{c|}{51.4} &
  \multicolumn{1}{c|}{3.7} &
  71.12 \\ \midrule
Llama3 8B &
  \multirow{5}{*}{XML} &
  \multicolumn{1}{c|}{12.9} &
  \multicolumn{1}{c|}{64.1} &
  \multicolumn{1}{c|}{6.1} &
  \multicolumn{1}{c|}{52.8} &
  \multicolumn{1}{c|}{\textbf{4.8}} &
  73.5 \\
Granite 8B &
   &
  \multicolumn{1}{c|}{3.6} &
  \multicolumn{1}{c|}{60.7} &
  \multicolumn{1}{c|}{2.8} &
  \multicolumn{1}{c|}{70.9} &
  \multicolumn{1}{c|}{10.7} &
  72.0 \\
Granite 20B &
   &
  \multicolumn{1}{c|}{2.1} &
  \multicolumn{1}{c|}{\textbf{53.3}} &
  \multicolumn{1}{c|}{1.9} &
  \multicolumn{1}{c|}{73.9} &
  \multicolumn{1}{c|}{12.2} &
  \textbf{70.5} \\
Granite 34B &
   &
  \multicolumn{1}{c|}{\textbf{1.9}} &
  \multicolumn{1}{c|}{56.9} &
  \multicolumn{1}{c|}{1.6} &
  \multicolumn{1}{c|}{63.1} &
  \multicolumn{1}{c|}{10.6} &
  71.9 \\
Codellama 34B &
   &
  \multicolumn{1}{c|}{2.3} &
  \multicolumn{1}{c|}{71.2} &
  \multicolumn{1}{c|}{1.6} &
  \multicolumn{1}{c|}{56.9} &
  \multicolumn{1}{c|}{10.2} &
  71.7 \\ \midrule
Llama3 8B &
  \multirow{5}{*}{YAML} &
  \multicolumn{1}{c|}{1.3} &
  \multicolumn{1}{c|}{53.3} &
  \multicolumn{1}{c|}{3.1} &
  \multicolumn{1}{c|}{62.4} &
  \multicolumn{1}{c|}{0.4} &
  74.5 \\
Granite 8B &
   &
  \multicolumn{1}{c|}{11.2} &
  \multicolumn{1}{c|}{\textbf{13.7}} &
  \multicolumn{1}{c|}{1.8} &
  \multicolumn{1}{c|}{63.9} &
  \multicolumn{1}{c|}{12.2} &
  70.5 \\
Granite 20B &
   &
  \multicolumn{1}{c|}{\textbf{1.6}} &
  \multicolumn{1}{c|}{39.8} &
  \multicolumn{1}{c|}{1.4} &
  \multicolumn{1}{c|}{56.6} &
  \multicolumn{1}{c|}{10.7} &
  72.0 \\
Granite 34B &
   &
  \multicolumn{1}{c|}{3.1} &
  \multicolumn{1}{c|}{14.9} &
  \multicolumn{1}{c|}{\textbf{1.1}} &
  \multicolumn{1}{c|}{\textbf{40.6}} &
  \multicolumn{1}{c|}{10.6} &
  71.9 \\
Codellama 34B &
   &
  \multicolumn{1}{c|}{7.1} &
  \multicolumn{1}{c|}{24.9} &
  \multicolumn{1}{c|}{1.4} &
  \multicolumn{1}{c|}{50.3} &
  \multicolumn{1}{c|}{12.6} &
  71.0 \\ \midrule
Llama3 8B &
  \multirow{5}{*}{Python} &
  \multicolumn{1}{c|}{5.4} &
  \multicolumn{1}{c|}{64.9} &
  \multicolumn{1}{c|}{3.1} &
  \multicolumn{1}{c|}{72.9} &
  \multicolumn{1}{c|}{\textbf{3.1}} &
  72.9 \\
Granite 8B &
   &
  \multicolumn{1}{c|}{2.4} &
  \multicolumn{1}{c|}{73.0} &
  \multicolumn{1}{c|}{2.3} &
  \multicolumn{1}{c|}{70.9} &
  \multicolumn{1}{c|}{10.7} &
  72.71 \\
Granite 20B &
   &
  \multicolumn{1}{c|}{\textbf{1.6}} &
  \multicolumn{1}{c|}{64.7} &
  \multicolumn{1}{c|}{2.4} &
  \multicolumn{1}{c|}{68.7} &
  \multicolumn{1}{c|}{16.6} &
  71.42 \\
Granite 34B &
   &
  \multicolumn{1}{c|}{2.6} &
  \multicolumn{1}{c|}{\textbf{61.2}} &
  \multicolumn{1}{c|}{\textbf{2.4}} &
  \multicolumn{1}{c|}{\textbf{66.9}} &
  \multicolumn{1}{c|}{8.9} &
  69.35 \\
Codellama 34B &
   &
  \multicolumn{1}{c|}{5.6} &
  \multicolumn{1}{c|}{65.1} &
  \multicolumn{1}{c|}{2.9} &
  \multicolumn{1}{c|}{64.1} &
  \multicolumn{1}{c|}{14.1} &
  \textbf{69.1} \\ \midrule
Llama3 8B &
  \multirow{5}{*}{NL} &
  \multicolumn{1}{c|}{5.8} &
  \multicolumn{1}{c|}{50.4} &
  \multicolumn{1}{c|}{3.4} &
  \multicolumn{1}{c|}{54.1} &
  \multicolumn{1}{c|}{5.6} &
  73.9 \\
Granite 8B &
   &
  \multicolumn{1}{c|}{\textbf{2.1}} &
  \multicolumn{1}{c|}{28.9} &
  \multicolumn{1}{c|}{2.6} &
  \multicolumn{1}{c|}{29.2} &
  \multicolumn{1}{c|}{8.3} &
  69.24 \\
Granite 20B &
   &
  \multicolumn{1}{c|}{2.9} &
  \multicolumn{1}{c|}{0.6} &
  \multicolumn{1}{c|}{2.8} &
  \multicolumn{1}{c|}{30.2} &
  \multicolumn{1}{c|}{7.97} &
  69.24 \\
Granite 34B &
   &
  \multicolumn{1}{c|}{2.3} &
  \multicolumn{1}{c|}{\textbf{1.9}} &
  \multicolumn{1}{c|}{\textbf{2.4}} &
  \multicolumn{1}{c|}{\textbf{8.9}} &
  \multicolumn{1}{c|}{9.86} &
  \textbf{63.42} \\
Codellama 34B &
   &
  \multicolumn{1}{c|}{2.8} &
  \multicolumn{1}{c|}{60.4} &
  \multicolumn{1}{c|}{2.9} &
  \multicolumn{1}{c|}{34.5} &
  \multicolumn{1}{c|}{\textbf{7.88}} &
  65.51 \\ \bottomrule
\end{tabularx}
\caption{Task 1 zero shot results having IS and RTV metric values. IS denotes the percentage of invalid samples and RTV denotes the percentage of sample root data type errors. For IS and RTV, the lesser the value better the performance.}

\label{table:task1isandrtvappendix}
\end{table*}

\begin{table}[h!]
\small
\centering
\begin{tabularx}{\columnwidth}{l|c|c}
\toprule
\textbf{Language} & \textbf{Max schema tokens} & \textbf{Avg schema tokens} \\
\midrule
XML & $3316$ & $364.82$ \\
JSON & $1954$ & $208.23$ \\
YAML & $1295$ & $135.09$ \\
\bottomrule
\end{tabularx}
\caption{Schema length comparison using Llama3 tokenizer}
\label{table:schema_length}
\end{table}

\subsection{Task Motivation} \label{sec:taskmotivation}

\subsubsection{Data as Code Generation Task}
This section describes use cases from enterprise and research points of view motivating data as code generation seed tasks in our study.

\paragraph{Enterprise Use Cases:} (i) Test case structured data generation to test application interfaces such as REST API endpoints. Often, enterprises have a large number of services exposing API endpoints that have to be tested, and LLMs can be a drop-in solution to generate test case data at scale. (ii) Structured configuration data generation for a particular use case and domain. Enterprise applications such as Kubernetes use DSLs for configuration and usage, preparing them require deep domain expertise and there is increasing motivation \cite{10247987} to employ LLMs in enterprises to generate DSL code. (iii) Some more downstream tasks involving structured data, such as forms and tables often represented in a programmable format such as JSON, can leverage LLMs to generate structured data to fill forms or tables leveraging the schema.

\paragraph{Research Use Cases:} (i) Since DSLs are typically low resource languages, LLMs are often employed \cite{song2020lightpafftwostagedistillationframework} to synthesize data from LLMs to train and evaluate smaller-sized models. (ii) This task acts a as a seed for similar NLP use cases such as code translation.

\begin{table*}[!ht]
\setlength\tabcolsep{24pt}
\scriptsize
\centering
\begin{tabularx}{\textwidth}{c|c|ccc}
\toprule
 &
   &
  \multicolumn{3}{c}{Output Representation} \\ \midrule
 &
   &
  \multicolumn{1}{c|}{JSON} &
  \multicolumn{1}{c|}{YAML} &
  \multicolumn{1}{c}{XML} \\ \midrule
Model &
  Schema &
  \multicolumn{1}{c|}{Macro-F1} &
  \multicolumn{1}{c|}{Macro-F1} &
  \multicolumn{1}{c}{Macro-F1} 
  \\ \midrule
Llama3 8B &
  \multirow{5}{*}{JSON} &
  \multicolumn{1}{c|}{0.55} &
  \multicolumn{1}{c|}{0.37} &
  \multicolumn{1}{c}{0.40} \\
Granite 8B &
   &
  \multicolumn{1}{c|}{0.55} &
  \multicolumn{1}{c|}{0.55} &
  \multicolumn{1}{c}{0.42} \\
Granite 20B &
   &
  \multicolumn{1}{c|}{0.48} &
  \multicolumn{1}{c|}{0.37} &
  \multicolumn{1}{c}{0.47} \\
Granite 34B &
   &
  \multicolumn{1}{c|}{0.60} &
  \multicolumn{1}{c|}{\textbf{0.56}} &
  \multicolumn{1}{c}{\textbf{0.63}} \\
Codellama 34B &
   &
  \multicolumn{1}{c|}{\textbf{0.64}} &
  \multicolumn{1}{c|}{0.53} &
  \multicolumn{1}{c}{0.50} \\ \midrule
Llama3 8B &
  \multirow{5}{*}{XML} &
  \multicolumn{1}{c|}{0.44} &
  \multicolumn{1}{c|}{0.35} &
  \multicolumn{1}{c}{0.41} \\
Granite 8B &
   &
  \multicolumn{1}{c|}{0.45} &
  \multicolumn{1}{c|}{0.44} &
  \multicolumn{1}{c}{0.50} \\
Granite 20B &
   &
  \multicolumn{1}{c|}{0.24} &
  \multicolumn{1}{c|}{0.45} &
  \multicolumn{1}{c}{\textbf{0.56}} \\
Granite 34B &
   &
  \multicolumn{1}{c|}{\textbf{0.52}} &
  \multicolumn{1}{c|}{\textbf{0.47}} &
  \multicolumn{1}{c}{0.39} \\
Codellama 34B &
   &
  \multicolumn{1}{c|}{0.41} &
  \multicolumn{1}{c|}{0.41} &
  \multicolumn{1}{c}{0.48} \\ \midrule
Llama3 8B &
  \multirow{5}{*}{YAML} &
  \multicolumn{1}{c|}{0.38} &
  \multicolumn{1}{c|}{0.40} &
  \multicolumn{1}{c}{0.40} \\
Granite 8B &
   &
  \multicolumn{1}{c|}{0.45} &
  \multicolumn{1}{c|}{0.50} &
  \multicolumn{1}{c}{0.44} \\
Granite 20B &
   &
  \multicolumn{1}{c|}{0.24} &
  \multicolumn{1}{c|}{0.31} &
  \multicolumn{1}{c}{0.45} \\
Granite 34B &
   &
  \multicolumn{1}{c|}{0.52} &
  \multicolumn{1}{c|}{\textbf{0.55}} &
  \multicolumn{1}{c}{0.47} \\
Codellama 34B &
   &
  \multicolumn{1}{c|}{\textbf{0.59}} &
  \multicolumn{1}{c|}{0.52} &
  \multicolumn{1}{c}{\textbf{0.58}} \\ \midrule
Llama3 8B &
  \multirow{5}{*}{Python} &
  \multicolumn{1}{c|}{0.37} &
  \multicolumn{1}{c|}{0.36} &
  \multicolumn{1}{c}{0.38} \\
Granite 8B &
   &
  \multicolumn{1}{c|}{0.54} &
  \multicolumn{1}{c|}{0.44} &
  \multicolumn{1}{c}{\textbf{0.54}} \\
Granite 20B &
   &
  \multicolumn{1}{c|}{0.34} &
  \multicolumn{1}{c|}{0.45} &
  \multicolumn{1}{c}{0.36} \\
Granite 34B &
   &
  \multicolumn{1}{c|}{\textbf{0.53}} &
  \multicolumn{1}{c|}{\textbf{0.47}} &
  \multicolumn{1}{c}{0.40} \\
Codellama 34B &
   &
  \multicolumn{1}{c|}{0.48} &
  \multicolumn{1}{c|}{0.45} &
  \multicolumn{1}{c}{0.46} \\ \midrule
Llama3 8B &
  \multirow{5}{*}{NL} &
  \multicolumn{1}{c|}{\textbf{0.63}} &
  \multicolumn{1}{c|}{\textbf{0.55}} &
  \multicolumn{1}{c}{0.57} \\
Granite 8B &
   &
  \multicolumn{1}{c|}{0.45} &
  \multicolumn{1}{c|}{0.51} &
  \multicolumn{1}{c}{0.39} \\
Granite 20B &
   &
  \multicolumn{1}{c|}{0.53} &
  \multicolumn{1}{c|}{0.45} &
  \multicolumn{1}{c}{\textbf{0.57}} \\
Granite 34B &
   &
  \multicolumn{1}{c|}{0.45} &
  \multicolumn{1}{c|}{0.46} &
  \multicolumn{1}{c}{0.38} \\
Codellama 34B &
   &
  \multicolumn{1}{c|}{0.52} &
  \multicolumn{1}{c|}{0.54} &
  \multicolumn{1}{c}{0.42} \\ \bottomrule
\end{tabularx}
\caption{Task 2 zero shot Macro-F1 scores. Task 2 is a binary classification task.}

\label{table:task2macrof1appendix}
\end{table*}

\subsubsection{DSL Validation Task}
This section describes use cases from an enterprise and research perspective that motivate our study's DSL validation seed task.

\paragraph{Enterprise Use Cases:} (i) Given the schema, employing LLMs to generate domain-aware suggestions over the provided structured data is not viable with traditional schema validators, which only pinpoint syntactic errors and cannot provide semantic suggestions. Such as providing optimizations over the existing resource YAML in Kubernetes while complying with resource schema. (ii) In an assistive chat system, the constraints are often in NL representation from the user, which is not machine-readable, and LLMs should be able to understand such constraints. (iii) Quick interoperability across different schema and data representation versions. Often in enterprises, schemas can be in a particular version that is incompatible with the structured data version. For instance, the schema could be in an older JSON schema version such as Draft 0 and data in Draft 7, in such cases LLMs can come handy to perform validation at scale.

\paragraph{Research Use Case:} Understanding LLMs' capability in validating the given structured data against the schema across representations can provide seed evidence for more complex tasks such as automatically fixing data in compliance with the given schema.

\subsection{Schema Examples} \label{sec:schemaex}
This section provides schemas across $5$ representations from \Cref{lst:samplejson,lst:sampleyaml,lst:samplepython,lst:samplexml,lst:samplenl}. All the schemas are equivalent in terms of constraints.
\begin{lstlisting}[language=json, caption=Sample schema using JSON Schema, label={lst:samplejson}, escapechar=@]
{
    "type": "object",
    "properties": {
        "footbaths": {
            "type": "boolean"
        },
        "deluded": {
            "type": "null"
        },
        "bravadoing": {
            "type": "number",
            "exclusiveMaximum": 5.131849487240756
        },
        "queintise": {},
        "manucodia": {
            "type": "number"
        },
        "antagonized": {},
        "outbacker": {
            "type": "number"
        },
        "sphenotripsy": {
            "type": "boolean"
        },
        "hw": {
            "type": "null"
        }
    },
    "additionalProperties": true,
    "required": []
}

\end{lstlisting}

\begin{lstlisting}[language=json, caption=Sample schema using YAML, label={lst:sampleyaml}, escapechar=@]
additionalProperties: true
properties:
  antagonized: {}
  bravadoing:
    exclusiveMaximum: 5.131849487240756
    type: number
  deluded:
    type: 'null'
  footbaths:
    type: boolean
  hw:
    type: 'null'
  manucodia:
    type: number
  outbacker:
    type: number
  queintise: {}
  sphenotripsy:
    type: boolean
required: []
type: object

\end{lstlisting}

\begin{lstlisting}[language=json, caption=Sample schema using Python, label={lst:samplepython}, escapechar=@]

from pydantic import BaseModel, Field

class Schema(BaseModel):
    footbaths: bool
    deluded: None = Field(None, alias="null")
    bravadoing: float = Field(..., exclusive_maximum=5.131849487240756)
    queintise: None = {}
    manucodia: float
    antagonized: None = {}
    outbacker: float
    sphenotripsy: bool
    hw: None = Field(None, alias="null")

\end{lstlisting}

\begin{lstlisting}[language=json, caption=Sample schema using XML, label={lst:samplexml}, escapechar=@]
<?xml version="1.0" ?>
<all>
	<type type="str">object</type>
	<properties type="dict">
		<footbaths type="dict">
			<type type="str">boolean</type>
		</footbaths>
		<deluded type="dict">
			<type type="str">null</type>
		</deluded>
		<bravadoing type="dict">
			<type type="str">number</type>
			<exclusiveMaximum type="float">5.131849487240756</exclusiveMaximum>
		</bravadoing>
		<queintise type="dict"/>
		<manucodia type="dict">
			<type type="str">number</type>
		</manucodia>
		<antagonized type="dict"/>
		<outbacker type="dict">
			<type type="str">number</type>
		</outbacker>
		<sphenotripsy type="dict">
			<type type="str">boolean</type>
		</sphenotripsy>
		<hw type="dict">
			<type type="str">null</type>
		</hw>
	</properties>
	<additionalProperties type="bool">true</additionalProperties>
	<required type="list"/>
</all>

\end{lstlisting}

\begin{lstlisting}[language=json, caption=Sample schema in NL , label={lst:samplenl}, escapechar=@]

This is a JSON schema that defines the structure of an object. Here's a breakdown of the schema:

# **Top-level properties**

# * `type`: The type of the JSON data, which is an object (`"object"`).
# * `properties`: An object that defines the properties of the object.
# * `additionalProperties`: A boolean value that indicates whether additional properties not specified in the schema are allowed. In this case, it is set to True
* required: An empty array that specifies no properties are required in the object.

**Properties object**

The `properties` object defines the structure of each property in the object. Here's a brief description of each property:

footbaths: A boolean
deluded: A null
bravadoing: A number that must be strictly lesser than 5.131849487240756, 
queintise: An object with no specific type or constraints.
manucodia: A number
antagonized: An object with no specific type or constraints.
outbacker: A number
sphenotripsy: A boolean
hw: A null

\end{lstlisting}

\end{document}